\documentclass[aps]{revtex4}

\usepackage{amsmath}
\usepackage{amssymb}

\newcommand{\be}{\begin{equation}}
\newcommand{\ee}{\end{equation}}

\newcommand{\bd}{\begin{displaymath}}
\newcommand{\ed}{\end{displaymath}}
\newcommand{\bea}{\begin{eqnarray}}
\newcommand{\eea}{\end{eqnarray}}

\begin{document}

\title{Quantum mechanics needs no interpretation}

\author{L. Sk\'ala$^{1,2}$\footnote{Corresponding author. 
E-mail: skala@karlov.mff.cuni.cz}
 and V. Kapsa$^1$}
\affiliation
{$^1$Charles University, Faculty of Mathematics and Physics, 
Ke Karlovu 3, 12116 Prague 2, Czech Republic} 
\affiliation
{$^2$University of Waterloo, Department of Applied Mathematics,\\
Waterloo, Ontario N2L 3G1, Canada}

\begin{abstract}
Probabilistic description of results of measurements and its consequences for 
understanding quantum mechanics are discussed.
It is shown that the basic mathematical structure of quantum mechanics
like the probability amplitude, Born rule, probability
density current, commutation relations, momentum operator, uncertainty 
relations, rules for including the scalar and vector
potentials and existence of antiparticles can be derived from
the definition of the mean values of the space coordinates and time.
Equations of motion of quantum mechanics, the Klein-Gordon equation,
Schr\"odinger equation and Dirac equation are obtained from requirement of
the relativistic invariance of the theory.
Limit case of localized probability densities leads to the
Hamilton-Jacobi equation of classical mechanics.
Many particle systems are also discussed.
\end{abstract}

\maketitle 

\noindent
keywords: probability theory, quantum mechanics, classical mechanics

\section{Introduction}

\label{introduction}

Quantum mechanics is one of the most thoroughly tested physical
theories (see e.g. \cite{Zeilinger2,Bertlmann,Scully1}).
At the same time, standard approach to introducing quantum mechanics based 
on the sometimes contra-intuitive postulates has rather mathematical than physical
character and exact physical meaning of the postulates
and their interpretation is subject of continuing discussion.
In this approach, quantum mechanics appears as a field with
strange paradoxes and phenomena that are not easy
to understand (see e.g. \cite{Laloe}).
It is not satisfactory and, in our opinion, it
is time (after almost 80 years after formulating the basic
principles of quantum mechanics) to replace this approach
by a more physical one based on a more direct description
of measurements.
Only such approach can clarify physical meaning of numerous
(sometimes hidden) assumptions made in quantum mechanics.
Paraphrasing the title of the recent paper by Fuchs and Peres \cite{Fuchs}
we can say: Quantum theory needs no `interpretation' -- it needs
derivation from description of measurements.
In this sense, our approach can be understood as 
extension and justification  
of the standard interpretation of quantum mechanics.

We note that our approach is different from that used usually in physics:
To explain experimental results, one introduces some physical quantities and
evolution equations these quantities have to obey. 
Then, consequences of these equations are investigated and 
compared with experiment.
In our approach, we describe results of measurements in a probabilistic way
and ask what is the mathematical apparatus that can describe this
situation. 
In this way, the basic mathematical structure of quantum theory
except for equations of motion is obtained. 
Equations of motion are found from requirement of the
relativistic invariance of the probabilistic description.

Probably the best approach is to start with measurement
of the space coordinates and time.
In this paper, we show that the basic mathematical structure of
quantum mechanics like the probability amplitude, Born rule, probability
density current, commutation relations, momentum operator, uncertainty 
relations, rules for including the scalar and vector
potentials and existence of antiparticles can be derived from
the definition of the mean values of the space coordinates and time
(sections \ref{uncertainty}-\ref{time}).
Equations of motion of quantum mechanics, the Klein-Gordon equation,
Schr\"odinger equation and Dirac equation are obtained from requirement of
the relativistic invariance of the theory (section \ref{equations}).
Limit case of localized probability densities yields the 
Hamilton-Jacobi equation of classical mechanics (section
\ref{classical}).
Generalization to many particle systems is performed in section \ref{many}.

\section{Uncertainty relation for probability density}

\label{uncertainty}
 
Physical measurements are imperfect and repeated measurements of
the same quantity under the same experimental conditions yield different results.
The most simple characteristics of such measurements
is the mean value of the results of repeated measurements.
It is the starting point of the following discussion.

We assume that the mean value $\langle x \rangle$ of the measurement 
of the coordinate $x$ is given by the usual definition    
\be
\label{caverx}
\langle x \rangle=\int x\rho({\bf r}){\rm d} V,
\ee
where the integration is carried out over the whole space and
$\rho({\bf r}) \geq 0$
is the normalized probability density
\be
\label{cnorm}
\int \rho\, {\rm d} V=1.
\ee
  
First, we use integration by parts with respect to the variable $x$ in
Eq. (\ref{cnorm}) and get
\be
\label{cpx0}
\left. x\rho\right|_{x=-\infty}^{\infty}-
\int x \frac{\partial\rho}{\partial x}{\rm d} V=1.
\ee
Assuming that the first term in this equation equals zero 
for physically reasonable $\rho$ we obtain
the starting point of the following discussion
\be
\label{cpx1}
\int x \frac{\partial\rho}{\partial x}{\rm d} V=-1.
\ee
The last equation
can be rewritten in form of the inner product
\be
\label{cuvx}
(u,v)=-1
\ee
defined in the usual way
\be
\label{cscalarx}
(u,v)=\int u^*({\bf r}) v({\bf r}) {\rm d} V.
\ee
Here, the star denotes the complex conjugate and functions $u$ and $v$
equal
\be
\label{cux}
u=x\sqrt{\rho},
\ee
\be
\label{cvx}
v=\frac{1}{\sqrt{\rho}}\frac{\partial\rho}{\partial x},
\ee
where $s=s({\bf r})$ is a real function.
Using the Schwarz inequality
\be
\label{Schwarz}
(u,u)(v,v) \geq |(u,v)|^2
\ee
and Eqs. (\ref{cuvx})-(\ref{cvx}) 
we can derive the uncertainty relation in the form 
\be
\label{cSchwarz1}
\int x^2 \rho\, {\rm d}V \int \frac{1}{\rho} 
\left(\frac{\partial\rho}{\partial x}\right)^2 {\rm d}V \geq 1,
\ee
where the second integral is the so-called Fisher information 
\cite{Fisher,Frieden4,Frieden,Cover}.
By using Eq. (\ref{cpx1}) and the condition $\rho\rightarrow 0$
for $x\rightarrow\pm\infty$ this result can be generalized to the  
Cramer-Rao inequality \cite{Frieden,Frieden6}
\be
\label{cSchwarz2}
\int (x-\langle x \rangle)^2 \rho\, {\rm d}V \int \frac{1}{\rho} 
\left(\frac{\partial\rho}{\partial x}\right)^2 {\rm d}V \geq 1.
\ee

In agreement with standard statistical description
of results of measurements, we require here the same weight function $\rho$
in the integrals $\int x\rho\,{\rm d}V$, $\int x^2\rho\,{\rm d}V$
and $\int (x-\langle x \rangle)^2\rho\,{\rm d}V$. 
Assumptions (\ref{cux}) and (\ref{cvx}) obey this requirement. 

{\em Uncertainty relation (\ref{cSchwarz2}) 
is general consequence of Eq. (\ref{caverx})
and shows that uncertainty relations must appear in any probabilistic
theory of this kind, including quantum mechanics.
There are two important quantities appearing in the
uncertainty relation (\ref{cSchwarz2}): the coordinate $x$ and the derivative 
$\partial/\partial x$.
Similar quantities, namely 
the coordinate $x$ and the momentum operator 
$-i\hbar\partial/\partial x$, 
appear also in quantum mechanics.}

\section{Born rule and complex probability amplitude}

\label{Born}

In the preceding section, we made use of the expression
$\rho=(\sqrt{\rho})^2$ in the inner product $(u,u)$ 
and derived the uncertainty relations (\ref{cSchwarz1}) and (\ref{cSchwarz2}).
However, we can assume also a general expression
\be
\label{apsi}
\rho=\psi^*\psi
\ee
where $\psi$ is a complex probability amplitude with the property 
\be
\label{psi1}
|\psi|=\sqrt{\rho}.
\ee
{\em Equation (\ref{apsi}) agrees with the well-known Born rule for
calculating the probability density from the probability amplitude.}
 
Physical meaning of the complex form of $\psi$ can be elucidated as follows.
To describe physical systems, we have to specify not only 
the space-time probability distribution $\rho$, 
but also its evolution in space-time.
This information can be encoded 
into the complex part of the probability amplitude $\psi$.
Here, we can proceed similarly as in continuum mechanics,
where not only the density  $\rho$ but also
the density current  
\be
\label{j_k}
j_k=\rho\, v_k, \quad k=1,2,3
\ee
related to the components of the velocity $v_k$ is introduced.
Writing the ``velocity'' in the form
\be
\label{v_k}
v_k= \frac{\partial s}{\partial x^k},
\ee
where $s=s({\bf r},t)$ is a real function we get
\be
\label{j1}
j_k=\rho\frac{\partial s}{\partial x^k}=
\sqrt{\rho}{\rm e}^{-is}\,\sqrt{\rho}{\rm e}^{is}
\frac{\partial s}{\partial x^k}
=\sqrt{\rho}{\rm e}^{-is}(-i)\frac{\partial(\sqrt{\rho}{\rm e}^{is})}{\partial x^k}
+\frac{i}{2}\frac{\partial \rho}{\partial x^k}.
\ee
Now, introducing the complex probability amplitude in agreement with
Eqs. (\ref{apsi})-(\ref{psi1}) 
\be
\label{psi}
\psi({\bf r},t)=\sqrt{\rho({\bf r},t)}{\rm e}^{is({\bf r},t)}
\ee
we have
\be
\label{j2}
j_k=\psi^*\left(-i\frac{\partial\psi}{\partial x^k}\right)
+\frac{i}{2}\frac{\partial \rho}{\partial x^k}.
\ee
However, the probability density current has to be real.
Calculating the real part of $j_k$ we obtain the final expression
\be
\label{j3}
j_k=
\frac{1}{2}\left[\psi^*\left(-i\frac{\partial \psi}{\partial
    x^k}\right)+c.c.\right]
=\frac{1}{2i}\left(\psi^*\frac{\partial \psi}{\partial
    x^k}-\psi\frac{\partial \psi^*}{\partial x^k}\right).
\ee
{\em Except for a multiplicative factor, this formula agrees
with the expression for the probability density current
known from quantum mechanics. 
The complex probability amplitudes $\psi$ are necessary to obtain
nonzero $j_k$.
The probability density current depends on the operator
$-i\partial/\partial x^k$.
Except for the factor $\hbar$, this operator agrees
with the momentum operator $-i\hbar\partial/\partial x^k$
known from quantum mechanics.
Therefore, this part of quantum mechanics can be derived
from general properties of the probability theory. 
In agreement with rules of quantum mechanics, 
the probability amplitudes $\psi$ and $\psi\exp(i\alpha)$, where $\alpha$
is a real constant, yield the same probability density $\rho$
and probability density current $j_k$.}

\section{Commutation relations for the coordinate and momentum}

Now we use Eqs. (\ref{cpx1}) and (\ref{apsi}) and get
\be
\label{basic}
\int x\left(\frac{\partial \psi^*}{\partial x}\psi
+\psi^*\frac{\partial \psi}{\partial x}\right)
{\rm d}V=-1.
\ee
Multiplying this equation by $-i$ we obtain the equation
\be
\label{apx2}
\int\left[(x\psi)^*\left(-i\frac{\partial\psi}{\partial x}\right)
-\left(-i\frac{\partial\psi}{\partial x}\right)^*x\psi\right]{\rm d}V=i
\ee
or
\be
\label{apx3}
[x, -i\partial/\partial x]=i.
\ee
{\em Except for the factor $\hbar\neq 0$, this commutation relation
agrees with the commutation relation between the coordinate and
momentum known from quantum mechanics.}

\section{Uncertainty relations for the coordinate and momentum}
  
Equation (\ref{basic}) can be rewritten as
\be
\label{un1}
2\,{\rm Re}(u,v)=-1,
\ee
where $u=x\psi$ and $v=\partial\psi/\partial x$.
Calculating the square of Eq. (\ref{un1}) we get successively
\be
\label{aineq}
1=4 [{\rm Re}(u,v)]^2 \leq 
\ee
\bd
\leq 4\{[{\rm Re}(u,v)]^2+[{\rm Im}(u,v)]^2\}=4|(u,v)|^2.
\ed
From this result and the Schwarz inequality (\ref{Schwarz})
we obtain the relation
\be
\label{auuvv}
(u,u)(v,v) \geq \frac{1}{4}
\ee
that can be written in form of the uncertainty relation 
\be
\label{auuvv1}
\int x^2 |\psi|^2\, {\rm d}V \int \left|-i\partial\psi/\partial x\right|^2 {\rm d}V \geq \frac{1}{4}.
\ee
This result can be further generalized.
Using the integration by parts and the condition $\rho\rightarrow 0$
for $x\rightarrow\pm\infty$ 
Eq. (\ref{basic}) can be generalized as
\be
\label{apx4}
\int [(x-a)\psi]^*
\bigg[\frac{\partial\psi}{\partial x}-i b\psi\bigg] {\rm d} V+
\ee
\bd
+\int\bigg[\frac{\partial\psi}{\partial x}-
i b\psi\bigg]^*[(x-a)\psi] {\rm d} V=-1,
\ed
where $a$ and $b$ are real constants.
From this equation, a more general form of the uncertainty relation 
can be obtained
\be
\label{auuvv2}
\int (x-a)^2 |\psi|^2\, {\rm d}V \int \left|-i\partial\psi/\partial x-b\psi\right|^2 {\rm d}V \geq \frac{1}{4}.
\ee
The minimum of the left side is obtained for
\be
\label{a}
a=\int \psi^* x \psi\, {\rm d} V=\langle x\rangle
\ee
and
\be
\label{b}
b=\int \psi^*(-i\partial\psi/\partial x)\, {\rm d} V=\langle -i\partial/\partial x \rangle.
\ee
{\em Except for the factor $\hbar$, the resulting uncertainty relation
with $a$ and $b$ given by the last two equations agrees 
with the well-known Heisenberg uncertainty relation.
Again, it follows from Eqs. (\ref{cpx1}) and (\ref{apsi}).}
 
It is worth noting that Eq. (\ref{apx4}) remains valid also in case 
if $b$ is a real function $b=f_x({\bf r},t)$.
It means that the operator $-i\partial\psi/\partial x$ can
be replaced by the operator $-i\partial\psi/\partial x-f_x$
and the commutation relation (\ref{apx3}) and
the uncertainty relation (\ref{auuvv2}) can be further generalized.
{\em Therefore, general structure of the probability
theory remains preserved for any real function $f_x$.
In physics, different functions $f_x$, $f_y$ and $f_z$ correspond to different 
components of the electromagnetic vector potential ${\bf A}=(A_x,A_y,A_z)$
multiplied by the charge $e$ of the particle.
Except for $\hbar$ and $e$, we obtained the rule 
$-i\hbar\nabla\rightarrow -i\hbar\nabla-e{\bf A}$
for including the vector potential ${\bf A}$ into quantum theory
(charge is discussed at the end of section \ref{time}).
We note also that the kinetic energy in quantum mechanics 
$T=(\hbar^2/2m)\int|\nabla\psi|^2{\rm d}V$ equals the space Fisher information
$\int|\nabla\rho|^2/\rho\,{\rm d}V$ multiplied by $\hbar^2/(8m)$. 
}

\section{Time}

\label{time}

Time can be discussed similarly as the space coordinates, however,
there are some important differences that has to be taken into consideration.

Assuming that there are given initial conditions for $\psi({\bf r},t=0)$  
the probability amplitude $\psi({\bf r},t)$, $t>0$ gives the probability description
of measurements at later times.
Therefore, time evolution has unidirectional character from given
initial conditions to the relative probability of results of (yet unperformed)
measurements at later times.
If this measurement is actually performed, the probabilistic description must
be replaced by a concrete result following from the performed
measurement.
It is the basis of two different evolution schemes in quantum
mechanics: time evolution described by the evolution equation
like the Schr\"odinger equation and 
reduction or collapse of the wave function.
In this paper, we are interested in the former case.
{\em Detailed description of the reduction of the probability amplitude
is not needed in our approach.}

In standard quantum mechanics, the probability amplitudes obey
the normalization condition $\int|\psi|^2{\rm d}V=1$ valid at all times
and the integral over time $\int_0^{\infty}\int |\psi|^2 {\rm d}V{\rm d}t$
goes to infinity.
This situation can be compared to that for a free particle.
For a free particle, the integral $\int|\psi|^2 {\rm d}V$ 
goes to infinity and $\psi$  
is usually normalized by means of the Dirac $\delta$-function.
For time, similar approach cannot be used for two reasons.
First, we do not perform here integration over all times,
but from the initial condition at $t=0$ to infinity.
Second, if the integral $\int_0^{\infty}\int|\psi|^2 {\rm d}V{\rm d}t$ 
goes to infinity we cannot 
define the mean time by analogy
with Eq. (\ref{caverx}) and proceed similarly as 
in the preceding sections.
For these reasons, we assume that not only the integral 
$\int|\psi|^2 {\rm d}V$ but also
the integral 
\be
\label{norm}
\int_0^{\infty}\int |\psi|^2 {\rm d}V{\rm d}t=1
\ee
equals one and proceed by analogy with the space coordinates.
In this way, we get the operator $i\partial/\partial t$, 
obtain the corresponding commutation and uncertainty relations 
and introduce the scalar potential.
At the end of our discussion, we will assume that 
$\int |\psi|^2 {\rm d}V$ changes negligibly in time,
normalize the probability amplitude by means of 
the usual condition $\int |\psi|^2 {\rm d}V=1$ and
perform transition to standard quantum mechanics.
 
First, we define the time component of the probability density
current by the equation analogous to Eqs. (\ref{j_k})-(\ref{v_k})
\be
\label{jt}
j_t=-\rho \frac{\partial s}{\partial t}
\ee
and obtain expression similar to Eq. (\ref{j3})
\be
j_t=\frac{1}{2}\left[\psi^*\left(i\frac{\partial \psi}{\partial t}\right)+c.c.\right].
\ee
{\em Except for a factor, this quantity equals the time component of the 
probability density current 
$j_0={\rm Re}[\psi^*i\hbar(\partial\psi/\partial x^0)]/m_0$ known
from relativistic quantum mechanics,
where $x^0=ct$ and $m_0$ is the rest mass.}
Then, by analogy with Eq. (\ref{apx2}) we derive the equation
\be
\label{apt2}
\int_{t=0}^{\infty}\int\left[\left(i\frac{\partial\psi}{\partial t}\right)^*t\psi
-(t\psi)^*\left(i\frac{\partial\psi}{\partial t}\right) \right]{\rm d}V{\rm d}t
=i.
\ee
One can introduce also a real constant $d$ into this equation
\be
\label{apt3}
\int_{t=0}^{\infty}\int
\left[\left(i\frac{\partial\psi}{\partial t}-d\psi\right)^*t\psi-\right.
\left.(t\psi)^*\left(i\frac{\partial\psi}{\partial t}-d\psi\right) \right]{\rm d}V{\rm d}t
=i.
\ee
The uncertainty relation for time can be written in form analogous
to Eq. (\ref{auuvv2})
\be
\label{tuuvv2}
\int_0^{\infty}\int t^2 |\psi|^2\,{\rm d}V{\rm d}t 
\int_0^{\infty}\int\left|i\partial\psi/\partial t-d\psi\right|^2 
{\rm d}V{\rm d}t \geq \frac{1}{4}.
\ee
Minimum of the left side is obtained for
\be
\label{bmin}
d=
\frac{1}{2}\left[\int_0^{\infty}\int \psi^*i(\partial\psi/\partial t) 
{\rm d}V{\rm d}t+c.c.\right].
\ee
Equation (\ref{tuuvv2}) is valid also if $d$
is replaced by a real function $f_0({\bf r},t)$.

To illustrate meaning of Eq. (\ref{tuuvv2}) we assume decaying probability amplitude
with the life time $\tau>0$
\be
\psi({\bf r},t)=
\frac{1}{\sqrt{\tau}}{\rm e}^{-i\omega t-t/(2\tau)}\psi({\bf r}),
\ee
where the space part of the probability amplitude is normalized
by the usual condition $\int|\psi({\bf r})|^2{\rm d}V=1$.
In this case, we get from Eqs. (\ref{tuuvv2})-(\ref{bmin})
\be
\int_0^{\infty}\int t^2 |\psi|^2\,{\rm d}V{\rm d}t=2\tau^2,
\ee
$d=\omega$ and
\be
\int_0^{\infty}\int\left|i\partial\psi/\partial t-d\psi\right|^2 
{\rm d}V{\rm d}t=\frac{1}{4\tau^2}.
\ee
Therefore, uncertainty relation (\ref{tuuvv2}) gives
the relation between the mean square time 
$\langle t^2\rangle=2\tau^2$
and the square of the imaginary part 
of the complex frequency $\omega-i/(2\tau)$
and has meaning of the time-energy uncertainty relation.   

In agreement with our understanding of direction of time, 
we assume that direct physical meaning have only 
the probability amplitudes corresponding to the non-negative values
of the time component of the probability density current integrated over
the whole space
\be
\label{j_t0}
\int j_t{\rm d}V=-\int\rho\frac{\partial s}{\partial t}{\rm d}V\ge 0.
\ee
If this quantity is negative, its sign can be reversed by 
the transformation $\psi\rightarrow\psi^*$ changing the sign
of the phase $s$ and the probability density currents $j_k$ and $j_t$. 
Performing this transformation we get from Eq. (\ref{apt3})
for $d=f_0$
\be
\label{apt7}
\int_{t=0}^{\infty}\int
\bigg[\bigg(i\frac{\partial\psi}{\partial t}+f_0\psi\bigg)^*t\psi-
(t\psi)^*\bigg(i\frac{\partial\psi}{\partial t}+f_0\psi\bigg) \bigg]
{\rm d}V{\rm d}t=i
\ee
and see that this transformation changes the sign of $f_0$.

Similar discussion can be done also for the space coordinates.
As a result, the transformation $\psi\rightarrow\psi^*$ leads
to change of sign of the functions $f_0$ and $f_k$, $k=1,2,3$
that can be respected by putting $f_0=eU$ and $f_k=eA_k$, where
$U$ and $A_k$ are the scalar and vector electromagnetic potentials.
{\em Therefore, the probability amplitudes $\psi$ 
and $\psi^*$ describe particles that differ by the sign of their 
charge and existence of particles and antiparticles agrees with general
structure of the probability theory and unidirectional
character of time.}

{\em Except for $\hbar$, we obtained also the rules 
$i\hbar\partial/\partial t\rightarrow i\hbar\partial/\partial t-eU$
and $-i\hbar\nabla\rightarrow -i\hbar\nabla-e{\bf A}$
for including the electromagnetic potentials into quantum theory.
These potentials representing different physical scenarios 
do not appear among the variables of the probability amplitude
and describe non-quantized classical fields.}

Now, we perform transition to standard quantum mechanics. 
In this limit case the integration over time need not be performed 
and the probability density can be normalized over the space only
$\int |\psi|^2 {\rm d}V=1$.
At the same time, the uncertainty relation (\ref{tuuvv2}) losts
its original meaning and time becomes a parameter rather than
a dynamical variable.
{\em It is the first reason for a different role of 
time and space coordinates in quantum mechanics.
The second reason is that the operator 
$i\partial/\partial t$ appears in equations of motion like
the Schr\"odinger equation and does not represent
an independent physical quantity.}
  
{\em It is worth noting that to obtain results of sections
\ref{uncertainty}-\ref{time}
no evolution equation has been needed.
Therefore, this part of the mathematical formalism of quantum mechanics
follows directly from the probabilistic description of results of measurements.
It is also interesting that the Planck constant
$\hbar$ does not appear in our discussion and 
can be included by multiplying Eqs. (\ref{apx2}) and
(\ref{apt2}) by $\hbar$.
Therefore, the Planck constant determines the units used in
measurements and scales at which
the probabilistic character of measurements is important.}

\section{Equations of motion}

\label{equations}

To find equations of motion we require
relativistic invariance of the theory.
Our approach is similar to that used
by Frieden who derives basic equations
of physics from the principle of extreme physical information \cite{Frieden}.  

First we note that all quantities discussed above depend on $\psi$
or its first derivatives with respect to time and space coordinates. 
Returning back to our scheme used in section \ref{time} we can create real 
relativistic invariant from the first derivatives of $\psi$ appearing 
in the uncertainty relations (\ref{auuvv2}) and (\ref{tuuvv2})
for $a=b=d=0$  
\be
\label{Kvv}
\int_0^{\infty}\int
\bigg(\frac{1}{c^2}\bigg|\frac{\partial\psi}{\partial t}\bigg|^2
-\sum_{k=1}^3\bigg|\frac{\partial\psi}{\partial x^k}\bigg|^2
\bigg){\rm d}V{\rm d}t={\rm const},
\ee
where $c$ is the speed of light.

Integral 
$\int_0^{\infty}\int|\partial\psi/\partial t|^2{\rm d}V{\rm d}t$
has meaning of the time Fisher information and is non-negative. 
Similar conclusion applies also for
$\int_0^{\infty}\int|\partial\psi/\partial x^k|^2{\rm d}V{\rm d}t$,
$k=1,2,3$. 
However, since Eq. (\ref{Kvv}) must be valid in all cases including
the case $\partial\psi/\partial x^k=0$ (in the language of
quantum mechanics, it corresponds to zero momentum and zero kinetic
energy) 
{\em we can conclude that ${\rm const}\ge 0$}.
  
In this equation, we can perform integration by parts with respect to
all variables.
For example, we get for time
\be
\label{pp}
\int_0^{\infty}\int
\frac{\partial\psi^*}{\partial t}\frac{\partial\psi}{\partial t}
{\rm d}V{\rm d}t=
\frac{1}{2}\bigg[\int
\bigg(\psi^*\frac{\partial\psi}{\partial t}
+c.c.\bigg){\rm d}V\bigg]_0^{\infty}-
\frac{1}{2}\int_0^{\infty}\int
\bigg(\psi^*\frac{\partial^2\psi}{\partial t^2}+c.c.
\bigg){\rm d}V{\rm d}t.
\ee
However, the first integral on the right side can be expressed as
$\partial(\int|\psi|^2{\rm d}V)/\partial t$ and disappears
in the limit of standard quantum mechanics when 
$\int|\psi|^2{\rm d}V=1$.
Analogous result can be obtained also for the variables $x^k$
assuming that $\partial|\psi|^2/\partial x^k$ for $x^k\rightarrow-\infty$
and $x^k\rightarrow\infty$ equal. 
In standard quantum mechanics, this condition is obeyed for a free
particle as well as for the bound states.
 
Now, we perform transition to standard quantum mechanics with the
wave function normalized in the usual way and get
\be
\label{Kvv1}
\frac{1}{2}\int\bigg[\psi^*
\bigg(\Delta-\frac{1}{c^2}\frac{\partial^2}{\partial t^2}
-{\rm const}\bigg)\psi+c.c.\bigg]{\rm d}V=0.
\ee
From here, we get equation of motion in the form
\be
\label{Klein}
\bigg(\Delta-\frac{1}{c^2}\frac{\partial^2}{\partial t^2}
-{\rm const}\bigg)\psi=0.
\ee
Since ${\rm const}\ge 0$, we can put ${\rm const}=m_0^2 c^2/\hbar^2$,
where $m_0$ is another constant, known as the rest mass of the particle.
{\em Therefore, requirement of the relativistic invariance applied
to quantities appearing in the probabilistic formulation leads to 
the Klein-Gordon equation for a free particle.}

The non-relativistic time Schr\"odinger equation
can be obtained from the Klein-Gordon equation by using the transformation
\be
\label{Spsi}
\psi={\rm e}^{m_0 c^2 t/(i\hbar)}\varphi,
\ee
where $\varphi$ is the probability amplitude appearing in the Schr\"odinger
equation.
This transition is known and will not be discussed here \cite{Davydov}.

The Dirac equation can be derived by replacing the probability
amplitude $\psi$ in Eq. (\ref{Kvv}) by a column vector with four components
\be
\label{KvvD0}
\int_0^{\infty}\int
\bigg(\frac{1}{c^2}\frac{\partial\psi^+}{\partial t}\frac{\partial\psi}{\partial t}
-\sum_{k=1}^3\frac{\partial\psi^+}{\partial x^k}\frac{\partial\psi}{\partial x^k}
\bigg){\rm d}V{\rm d}t={\rm const},
\ee
where the cross denotes the hermitian conjugate.
Inserting the $\gamma^{\mu}$ matrices \cite{Davydov}
into this equation, putting ${\rm const}=m_0^2c^2/\hbar^2$ 
and using Eq. (\ref{norm}) we get
\be
\label{KvvD}
\int_0^{\infty}\int
\bigg[\frac{1}{c^2}
\bigg(\gamma^0\frac{\partial\psi}{\partial t}\bigg)^+
\bigg(\gamma^0\frac{\partial\psi}{\partial t}\bigg)-
\sum_{k=1}^3
\bigg(\gamma^k\frac{\partial\psi}{\partial x^k}\bigg)^+
\bigg(\gamma^k\frac{\partial\psi}{\partial x^k}\bigg)
-\frac{m_0^2 c^2}{\hbar^2}\psi^+\psi\bigg]{\rm d}V{\rm d}t=0.
\ee
Then, using properties of the $\gamma^{\mu}$ matrices and 
assuming that the integration by parts can be used analogously 
as in case of Eq. (\ref{pp}) the last equation leads in the limit of standard
quantum mechanics to
(see also \cite{Frieden4,Frieden})
\be
\label{Dvv3}
\int\bigg(\frac{\gamma^0}{c}\frac{\partial\psi}{\partial t}
-\sum_{k=1}^3 \gamma^k\frac{\partial\psi}{\partial x^k}
-\frac{im_0 c}{\hbar}\psi\bigg)^+
\bigg(\frac{\gamma^0}{c}\frac{\partial\psi}{\partial t}
+\sum_{k=1}^3\gamma^k\frac{\partial\psi}{\partial x^k} 
+\frac{im_0 c}{\hbar}\psi\bigg){\rm d} V=0.
\ee
The operator in the first parentheses is the hermitian
conjugate of that in the second ones.
Assuming that the expression in the second parentheses equals zero 
we obtain the Dirac equation for a free particle
\be
\label{Diraceq}
\frac{\gamma^0}{c}\frac{\partial\psi}{\partial t}
+\sum_{k=1}^3\gamma^k\frac{\partial\psi}{\partial x^k}+
\frac{im_0c}{\hbar}\psi=0.
\ee
{\em We see that requirement of the relativistic invariance
of the probabilistic description yields 
all the basic equations of motion of quantum mechanics.
The scalar and vector potentials can be included by means of the rules
$i\hbar\partial/\partial t\rightarrow i\hbar\partial/\partial t-eU$
and 
$-i\hbar\nabla\rightarrow -i\hbar\nabla-e{\bf A}$ discussed above.
}

\section{Classical mechanics}

\label{classical}

To derive the Hamilton-Jacobi equation 
for a free particle we proceed as follows. 
The probability amplitude is assumed in the form
\be
\label{rpsi}
\psi={\rm e}^{is/\hbar}={\rm e}^{is_1/\hbar}{\rm e}^{-s_2/\hbar},
\ee
where $s_1$ and $s_2$ are the real and imaginary parts of $s$,
respectively.
In the limit of standard quantum mechanics mentioned above, 
Eq. (\ref{Kvv}) with ${\rm const}=m_0^2c^2/\hbar^2$ can
be replaced by the equation
\be
\label{rHJ}
\frac{1}{c^2}\int\bigg|\frac{\partial s}{\partial t}\bigg|^2 
|\psi|^2 {\rm d} V=\int\big|\nabla s\big|^2|\psi|^2 {\rm d} V+m_0^2c^2.
\ee
Now we assume that the probability density 
\be
\label{rrho}
\rho=|\psi|^2={\rm e}^{-2s_2/\hbar}
\ee 
has very small values everywhere except for the vicinity of the point 
$\langle{\bf r}\rangle$, 
where it achieves its maximum and the first derivatives of $s_2$ 
at this point equal zero  
\be
\label{rder}
\left. \frac{\partial s_2}{\partial x^k}\right|_{{\bf r}=
\langle{\bf r}\rangle} =0,\quad k=1,2,3.
\ee
In such a case, the probability density can be replaced by
the $\delta$-function
\be
\label{rdelta}
|\psi|^2=
\delta({\bf r}-\langle {\bf r}\rangle)
\ee
and probabilistic character of the theory disappears.
Equations (\ref{rHJ})-(\ref{rdelta}) then lead to the relativistic 
equation
\be
\label{rHJcl}
\frac{1}{c^2}
\bigg(\frac{\partial s_1(\langle {\bf r}\rangle,
\langle t\rangle)}{\partial t}\bigg)^2=
[\nabla s_1(\langle {\bf r}\rangle,t)]^2+m_0^2c^2.
\ee
We note that Eq. (\ref{rdelta}) corresponds to the limit 
$\hbar\rightarrow 0$ in Eq. (\ref{rrho}).
Therefore, $s_1$ in Eq. (\ref{rHJcl}) is in fact the first term 
of the expansion of $s_1$ into the power series in $\hbar$
\be
s_1=s_1|_{\hbar=0}.
\ee
Further, 
we replace the mean coordinates $\langle{\bf r}\rangle$
by ${\bf r}$ as it is
usual in classical mechanics and 
introduce the classical non-relativistic action $S({\bf r},t)$
\be
\label{raction}
s_1=S-m_0c^2t.
\ee 
Equation (\ref{rHJcl}) then leads to 
\be
\label{rHJcl0}
\frac{1}{c^2}\bigg(\frac{\partial S}{\partial t}-m_0c^2\bigg)^2=
(\nabla S)^2+m_0^2c^2.
\ee 
In the non-relativistic limit
$|\partial S/\partial t|\ll m_0c^2$
the last equation yields 
the Hamilton-Jacobi equation for a free particle
\be
\label{rHJcl1}
\frac{\partial S}{\partial t}+\frac{(\nabla S)^2}{2m_0}=0.
\ee 
{\em Thus, the Hamilton-Jacobi equation can be obtained
from the probabilistic description of measurements
in the limit of $\delta$-like probability densities and 
non-relativistic approximation.
The scalar and vector potentials $U$ and ${\bf A}$ 
can be included by means of the rules 
$\partial S/\partial t\rightarrow \partial S/\partial t+eU$
and
$\nabla S\rightarrow \nabla S-e{\bf A}$
following from the rules 
$i\hbar\partial/\partial t\rightarrow i\hbar\partial/\partial t-eU$
and 
$-i\hbar\nabla\rightarrow -i\hbar\nabla-e{\bf A}$ discussed above.}

\section{Many particle systems}

\label{many}

In general, many particle systems have to be described by
the quantum field theory.
However, if we limit ourselves to 
quantum mechanics, we can proceed as follows. 

The starting point of discussion of the $N$ particle system is 
definition analogous to Eq. (\ref{caverx})  
\be
\label{caverxN}
\langle {\bf r}_j \rangle=
\int {\bf r}_j\rho({\bf r}_1,\ldots,{\bf r}_N,t){\rm d} V_1\ldots{\rm d}V_N,
\;j=1,\ldots,N,
\ee
where $\rho$ is the many particle probability density
and ${\bf r}_j$ are the coordinates of the $j$-th particle.
Then, discussion can be performed analogously to that given
above and the probability amplitude, uncertainty and commutation
relations, momentum operators and density currents for all
particles can be introduced. 
The scalar and vector potentials $U({\bf r}_1,\ldots,{\bf r}_N,t)$ and
${\bf A}({\bf r}_1,\ldots,{\bf r}_N,t)$
and antiparticles can be also discussed.  

Equations of motion for $N$ free particles can be found from 
generalization of the relativistic invariant (\ref{Kvv})  
\be
\label{KvvN}
\int_0^{\infty}\int
\bigg(\frac{1}{c^2}\bigg|\frac{\partial\psi}{\partial t}\bigg|^2
-\sum_{j=1}^N|\nabla_j\psi|^2
\bigg){\rm d}V_1\ldots{\rm d}V_N{\rm d}t=
\sum_{j=1}^N\frac{m_j^2c^2}{\hbar^2},
\ee
where $\psi({\bf r}_1,\ldots,{\bf r}_N,t)$ is the $N$ particle
probability amplitude
and $m_j$ denotes the rest mass of the particle.

Using similar approach as above, we can then obtain the 
$N$ particle Schr\"odinger equation
\be
\label{SchrodN}
-\sum_{j=1}^N\frac{\hbar^2}{2m_j}\Delta_j\psi=i\hbar\frac{\partial\psi}{\partial t}
\ee
and the Hamilton-Jacobi equation
\be
\label{rHJclN}
\frac{\partial S}{\partial t}+\sum_{j=1}^N\frac{(\nabla_j S)^2}{2m_j}=0.
\ee

For a system of identical particles, the probability density $\rho$ 
must be symmetric with respect to the exchange
of any two particles $i$ and $j$.
Hence, the probability amplitude $\psi$ must be
symmetric or antisymmetric with respect to such exchanges.

Non-locality of quantum mechanics is related to the many 
particle character of the probability density 
$\rho$ and the corresponding
probability amplitude $\psi$.

{\em It is seen that probabilistic description of measurements 
and its relativistic invariance yields also the basic
mathematical structure of the many particle quantum mechanics.}

\section{Conclusions}

In this paper, we have shown that the basic mathematical structure
of quantum mechanics can be derived from the probabilistic description of 
the results of measurement of the space coordinates and time. 
Equations of motion of quantum mechanics have been obtained from
requirement of the relativistic invariance of the theory.
As a limit case, this approach yields also the Hamilton-Jacobi
equation of classical mechanics. 
 
Since our approach makes possible to obtain the most significant parts 
of the mathematical formalism of quantum mechanics from
the probabilistic description of results of measurements, 
we believe that it is natural and 
physically satisfactory starting point to understanding
this field.
It contributes also to understanding quantum theory as correctly
formulated probabilistic description of measurements 
that can describe physical phenomena        
at different levels of accuracy from the most simple models to 
very complex ones.

\acknowledgements{This work was supported by
GA CR (grant No. 202/03/0799) and MS (grant No. 190-01/206053)
of the Czech Republic.}

\end{document}